\documentclass[prd,twocolumn,showpacs,preprintnumbers,superscriptaddress,ams 
math]{revtex4}
\begin{document}
\author{Patricio Salgado}
\altaffiliation{pasalgad@udec.cl} \affiliation{Departamento de
F\'{\i}sica, Universidad de Concepci\'{o}n,\\ Casilla 160-C,
Concepci\'{o}n, Chile. \\ Max Planck Institut f\"{u}r Gravitationsphysik, 
Albert Einstein\\
Institut, Am M\"{u}hlenberg 1, D-14476 Golm bei Potsdam, Germany.}
\author{Sergio del Campo}
\altaffiliation{sdelcamp@ucv.cl} \affiliation{Instituto de
F\'{\i}sica, Pontificia Universidad Cat\'{o}lica de Valpara\'{\i}so,\\
Avda Brasil 2950, Valpara\'{\i}so, Chile.}

\author{Mauricio Cataldo}
\altaffiliation{mcataldo@ubiobio.cl} \affiliation{Departamento de
F\'\i sica, Facultad de Ciencias, Universidad del B\'\i o-B\'\i o,
Avenida Collao 1202, Casilla 5-C, Concepci\'on, Chile.}

\title{${\bf N=1}$ Super{\bf gravity with cosmological constant and the 
}$AdS${\bf
\ group}}
\begin{abstract}
It is shown that the supersymmetric extension of the Stelle-West
formalism permits the construction of an action for
$(3+1)$-dimensional $N=1$ supergravity with cosmological constant
genuinely invariant under the $OSp(4/1).$ Since the action is
invariant under the supersymmetric extension of the $AdS$ group,
the supersymmetry algebra closes off shell without the need for
auxiliary fields. The limit case $m\rightarrow 0$, i.e.$(3+1)$
-dimensional $N=1$ supergravity invariant under the Poincar\'{e}
supergroup is also discussed. PACS number(s): 04.65. +e
\end{abstract}

\maketitle
\section{\bf Introduction}

In recent years it has been shown that in odd-dimensional supergravities
\cite{ban}, \cite{zan}: the fundamental field is always the connection $A$
and, in their simplest form, these are pure Chern-Simons systems. In
contrast with the standard cases, the supersymmetry transformations close
off-shell without auxiliary fields.

The Chern-Simons construction fails in even-dimensions for the simple reason
that there has not been found a characteristic class constructed with
products of curvature in odd dimensions. This could be a reason why the
construction of a (super)gravity in even dimensions invariant under the
(anti) de Sitter group has remained as an interesting open problem.

It is the purpose of this paper to show that the supersymmetric
extension of the Stelle-West formalism \cite{stelle}, which is an
application of the theory of nonlinear realizations to gravity,
permits constructing a $(3+1)$ -dimensional supergravity off-shell
invariant under the (anti) de Sitter group. Applications of the
theory of nonlinear realizations to supergravity has been carried
out by Chang and Mansouri \cite{mansouri}, and by G\"{u}rsey and
Marchildon \cite{gursey}. These authors considered a nonlinear
realization of the $OSP(1,4)$ in the context of the spontaneous
breakdown of supergravity. Unlike the present work, they
identified the corresponding coset parameters with the points of
space-time itself.

In the present work, the Goldstone fields represent a point in an internal
anti-de Sitter space. In describing the geometry of this internal space, we
make use of some of the results of ref. \cite{zumino} on the nonlinear
realization of supersymmetry in anti-de Sitter space.

An important stimulus for the interest in the construction of a supergravity
invariant under the $AdS$ superalgebra has come from recent developments in
M theory \cite{Witt}. In particular, Some of the expected features of
M-theory are (i) its dynamics should somehow exhibit a superalgebra in which
the anticommutator of two supersymmetry generators coincides with the $AdS$
superalgebra in eleven dimensions \cite{town}, (ii) the low-energy regime
should be described by an eleven dimensional supergravity of a new type
which should stand on a firm geometric foundation in order to have an
off-shell local supersymmetry \cite{nish}

The paper is organized as follows: In sec.$II$, we shall review some aspects
of the torsion-free condition in supergravity with cosmological constant.
The Supersymmetric extension of the Stelle-West formalism is carried out in
sec.$III$ where the principal features of the nonlinear realizations are
reviewed and the nonlinear fields vierbein, spin connection and gravitino
are derived. An action for supergravity genuinely invariant under the $AdS$
superalgebra is constructed in sec. $IV$, and its corresponding field
equations as well as the limit $m\rightarrow 0$ are discussed. Section $V$
concludes the work with a look forward to applications of the present
results to supergravity in higher dimensions. Some technical details on the
calculations are presented in the Appendix.

\section{{\bf \ }$N=1${\bf \ supergravity}}

In this section we shall review some aspects of the torsion-free condition
in supergravity.

\subsection{\bf The torsion-free condition in N=1 supergravity}

Supergravity is the theory of the gravitational field interacting with a
spin $3/2$ Rarita Schwinger field \cite{Des}, \cite{Freed}, \cite{van}. In
the simplest case there is just one spin $3/2$ Majorana fermion, usually
called the gravitino $\psi $. The corresponding action is

\begin{equation}
S=\int \varepsilon _{abcd}e^ae^bR^{cd}+4\overline{\psi }\gamma _5e^a\gamma
_aD\psi   \label{su0}
\end{equation}
where, $e^a$ is the 1-form vielbein, $\omega ^{ab}$ is the 1-form spin
connection, and $D\psi =d\psi -\frac 12$ $\omega ^{ab}\gamma _{ab}\psi $ is
the Lorentz covariant derivative.

$D=3+1$, $N=1$ supergravity is based on the Poincar\'{e} supergroup whose
generators $P_a,J_{ab},Q^\alpha $ satisfy the following Lie-superalgebra:
\[
\left[ P_a,P_b\right] =0
\]
\[
\left[ J_{ab},P_c\right] =i\left( \eta _{ac}P_b-\eta _{bc}P_a\right)
\]
\[
\left[ J_{ab},J_{cd}\right] =i\left( \eta _{ac}J_{bd}-\eta _{bc}J_{ad}+\eta
_{bd}J_{ac}-\eta _{ad}J_{bc}\right)
\]
\[
\left[ J_{ab},Q_\alpha \right] =i\left( \gamma _{ab}\right) _{\alpha \beta
}Q_\beta
\]
\[
\left[ P_a,Q_\beta \right] =0
\]
\begin{equation}
\left[ Q_\alpha ,\overline{Q}_\beta \right] =-2\left( \gamma ^a\right)
_{\alpha \beta }P_a.  \label{su1}
\end{equation}

Working in first order formalism, the gauge fields $e^a$, $\omega
^{ab}$, $ \psi $ are treated as independent. The key observation
is that $(e^a,\omega ^{ab},\psi )$, considered as a single entity,
constitute a multiplet in the adjoint representation of the
Poincare supergroup. That is, we can write:

\begin{equation}
A=A^AT_A=\frac 12i\omega ^{ab}J_{ab}-ie^aP_a+\overline{\psi }Q  \label{su2}
\end{equation}
where $A$ is the gauge field of the Poincar\'{e} supergroup, $
P_a,J_{ab},Q^\alpha $ being the generators of the Poincare
translations, of the Lorentz transformations and of the
supersymmetry, respectively. Hence supergravity is the gauge
theory of the Poincar\'{e} supergroup.

The field strength associated with $A^A$ is defined as the Poincar\'{e} Lie
superalgebra-valued curvature $2$-form $R^A.$ Splitting the index $A$, we
get
\begin{equation}
R^{ab}=d\omega ^{ab}-\omega _c^a\omega ^{cd}  \label{su8}
\end{equation}

\begin{equation}
\stackrel{\wedge }{T\text{ }}^a=T\ ^a-\frac 12\stackrel{\_}{\psi }\gamma
^a\psi   \label{su9}
\end{equation}
\begin{equation}
\rho =D\psi .  \label{su10}
\end{equation}

The associated Bianchi identities are given by
\begin{equation}
DR^{ab}=0  \label{su8'}
\end{equation}
\begin{equation}
DT\ ^a+R^{ab}e_b-i\stackrel{\_}{\psi }\gamma ^a\rho =0  \label{su9'}
\end{equation}
\begin{equation}
D\rho +\frac 14R^{ab}\gamma _{ab}\psi =0.  \label{su10'}
\end{equation}

However, although $A^A\equiv (e^a,\omega ^{ab},\psi )$ is a Yang\_Mills
potential and $R^A\equiv (R^{ab},\stackrel{\wedge }{T\text{ }}^a,\rho )$ the
corresponding field strength, the action (\ref{su0}) is not of the
Yang-Mills type. The main differences between an action of the Yang-Mills
type and the action (\ref{su0})are: (i) a Yang-Mills action is invariant
under the whole gauge group of which the $A^A$ are the Lie superalgebra
valued potentials; (ii) the action (\ref{su0}), instead, is not invariant
under the whole gauge supergroup, but only under the Lorentz transformations.

The invariance under Lorentz gauge transformations is manifest. To show the
non invariance of (\ref{su0}) both under  a supergauge translation and under
supersymmetry we recall that, under any gauge transformation, the gauge
connection $A^A$ transforms as

\begin{equation}
\delta A=-D\lambda =d\lambda -\left[ A,\lambda \right]   \label{su3}
\end{equation}
with
\begin{equation}
\lambda =\frac 12i\kappa ^{ab}J_{ab}-i\rho ^aP_a+\overline{\varepsilon }Q
\label{su4}
\end{equation}
Using the algebra (\ref{su1}) we obtain that $e^a$, $\omega
^{ab},$ and $ \psi $, under Poincar\`{e} translations, transform
as
\begin{equation}
\delta e^a=D\rho ^a;\quad \delta \omega ^{ab}=0\text{; \quad }\delta \psi =0;
\label{su5}
\end{equation}
under Lorentz rotations, as
\begin{equation}
\delta e^a=\kappa _b^ae^b;\quad \delta \omega ^{ab}=D\kappa ^{ab};\quad
\delta \psi =-\frac 12\kappa ^{ab}\gamma _{ab}\psi ;  \label{su6}
\end{equation}
and under supersymmetry transformations, as
\begin{equation}
\delta e^a=-2i\overline{\varepsilon }\gamma ^a\psi ;\quad \delta \omega
^{ab}=0\text{; \quad }\delta \psi =D\varepsilon .  \label{su7}
\end{equation}

The action (\ref{su0}) is invariant under diffeomorphism, and under local
Lorentz rotations , but it is not invariant under neither Poincare
translations nor supersymmetry.

In fact, under local Poincare translations
\[
\delta S_{pt}=2\int \varepsilon _{abcd}R^{ab}\left( T^c-\frac
12\overline{ \psi }\gamma ^c\psi \right) \rho ^d+\text{surf. term}
\]
\begin{equation}
\delta S=2\int \varepsilon _{abcd}R^{ab}\stackrel{\wedge
}{T}^c\rho ^d+\text{ surf}.\text{ term. }  \label{su12}
\end{equation}

Under local supersymmetry transformations
\begin{equation}
\delta S_{susy}=-4\int \overline{\varepsilon }\gamma _5\gamma _aD\psi
\stackrel{\wedge }{T\text{ }}^a+\text{surf. term.}  \label{su13}
\end{equation}
Thus the invariance of the action requires the vanishing of the torsion
\begin{equation}
\stackrel{\wedge }{T\text{ }}^a=0.  \label{su14}
\end{equation}
This means that the connection is no longer an independent variable. Rather,
its variation is given in terms of $\delta e^a$ and $\delta \psi $, and
differs from the one dictated by group theory. An effect of the
supertorsion-free condition on the local Poincar\'{e} superalgebra is that
all commutators on $e^a,$ $\psi $ close except the commutator of two local
supersymmetry transformations on the gravitino. For this commutator on the
vierbein one finds
\begin{equation}
\left[ \delta \left( \varepsilon _1\right) ,\delta \left( \varepsilon
_2\right) \right] e^a=\frac 12\overline{\varepsilon }_2\gamma ^aD\varepsilon
_1-\frac 12\overline{\varepsilon }_1\gamma ^aD\varepsilon _2=\frac 12D\left(
\overline{\varepsilon }_2\gamma ^a\varepsilon _1\right) .  \label{su15}
\end{equation}
With $\rho ^a=\frac 12\overline{\varepsilon }_2\gamma ^a\varepsilon _1,$ we
can write
\begin{equation}
\left[ \delta \left( \varepsilon _1\right) ,\delta \left( \varepsilon
_2\right) \right] e^a=D\rho ^a.  \label{su16}
\end{equation}
This means that, in the absence of the torsion-free condition, the
commutator of two local supersymmetry transformations on the vierbein is a
local Poincar\'{e} translation. However, the action is invariant by
constructi\'{o}n under general coordinate transformations, but not under
local Poincar\'{e} translation. The general coordinate transformation and
the local Poincar\'{e} translation can be identified if we impose the
torsion-free condition: since $\rho ^a=\rho ^\nu e_\nu ^a$ we can write
\[
D_\mu \rho ^a=\left( \partial _\mu \rho ^\nu \right) e_\nu ^a+\rho
^\nu \left( \partial _\nu e_\mu ^a\right) +\frac 12\rho ^\nu
\left( \stackrel{\_}{ \psi }_\mu \gamma ^a\psi _\nu \right)
\]
\begin{equation}
+\rho ^\nu \omega _\nu ^{ab}e_{\mu b}+\rho ^\nu T_{\mu \nu }^a.
\end{equation}
This means that, if $T_{\mu \nu }^a=0,$ then the following commutator is
valid:

\[
\left[ \delta _Q\left( \varepsilon _1\right) ,\delta _Q\left( \varepsilon
_2\right) \right] =\delta _{GCT}\left( \rho ^\mu \right) +\delta
_{LLT}\left( \rho ^\mu \omega _\mu ^{ab}\right)
\]
\begin{equation}
+\delta _Q\left( \rho ^\nu \stackrel{\_}{\psi }_\nu \right)  \label{su17}
\end{equation}

where we can see that $P$ in $\left\{ Q,Q\right\} =P,$ i.e. local Poincar\'e
translation, is replaced by general coordinate transformations besides two
other gauge symmetries. The structure constants defined by this result are
field-dependent \cite{van}, which is a property of supergravity not present
in Yang-Mills Theory .

The commutator of two local supersymmetry transformations on the gravitino
is given by
\[
\left[ \delta \left( \varepsilon _1\right) ,\delta \left( \varepsilon
_2\right) \right] \psi =\frac 12\left( \sigma _{ab}\varepsilon _2\right)
\left[ \delta \left( \varepsilon _1\right) \omega ^{ab}\right]
\]
\begin{equation}
-\frac 12\left( \sigma _{ab}\varepsilon _1\right) \left[ \delta \left(
\varepsilon _2\right) \omega ^{ab}\right] .  \label{su18}
\end{equation}

The condition $\stackrel{\wedge }{T}^a=0$ leads to $\omega
^{ab}=\omega ^{ab}(e,\psi )$ which implies that the connection is
no longer an independent variable, and its variation $\delta
\left( \varepsilon \right) \omega ^{ab}$ is given in terms of
$\delta \left( \varepsilon \right) e^a$ and $\delta \left(
\varepsilon \right) \psi .$ Introducing $\delta \left( \varepsilon
\right) \omega ^{ab}(e,\psi )$ into (\ref{su18}) we see that,
without the auxiliary fields, the gauge algebra does not close, as
shows the eq. $\left( 10\right) $ of ref.\cite{van}. Therefore the
condition $ \stackrel{\wedge }{T}^a=0$ not only breaks local
Poincare invariance, but also the supersymmetry transformations.

\subsection{{\bf The torsion-free condition in }$N=1${\bf \ supergravity
with cosmological constant}}

The action for supergravity with cosmological constant is given by \cite
{townsend}

\[
S=\int \varepsilon _{abcd}R^{ab}e^ce^d+4\overline{\psi }\gamma _5\gamma
_aD\psi e^a
\]
\begin{equation}
+2\alpha ^2\varepsilon _{abcd}e^ae^be^ce^d+3\alpha \varepsilon
_{abcd} \overline{\psi }\gamma ^{ab}\psi e^ce^d  \label{ssu0}
\end{equation}
where, $e^a$ is the 1-form vielbein, $\omega ^{ab}$ is the 1-form spin
connection, and $D\psi =d\psi -\frac 12$ $\omega ^{ab}\gamma _{ab}\psi $ is
the Lorentz covariant derivative.

The anti de Sitter version of $N=1,$ $D=3+1$ supergravity is based
on the graded extension of the $AdS$ group, i.e on the $OSp(1/4)$
whose generators $ P_a,J_{ab},Q^\alpha $ satisfy the following
Lie-superalgebra:
\[
\left[ P_a,P_b\right] =-im^2J_{ab}
\]
\[
\left[ J_{ab},P_c\right] =i\left( \eta _{ac}P_b-\eta _{bc}P_a\right)
\]
\[
\left[ J_{ab},J_{cd}\right] =i\left( \eta _{ac}J_{bd}-\eta _{bc}J_{ad}+\eta
_{bd}J_{ac}-\eta _{ad}J_{bc}\right)
\]
\[
\left[ J_{ab},Q_\alpha \right] =i\left( \gamma _{ab}\right) _{\alpha \beta
}Q_\beta
\]
\[
\left[ P_a,Q_\alpha \right] =-\frac i2m(\gamma _a)_{\alpha \beta }Q_\beta
\]
\begin{equation}
\left[ Q_\alpha ,\overline{Q}_\beta \right] =-2\left( \gamma ^a\right)
_{\alpha \beta }P_a-2m(\gamma ^{ab})_{\alpha \beta }J_{ab}.  \label{ssu1}
\end{equation}

Working in first order formalism, the gauge fields $e^a$, $\omega
^{ab}$, $ \psi $ are treated as independent. The key observation
is that $(e^a,\omega ^{ab},\psi )$, considered as a single entity,
constitute a multiplet in the adjoint representation of the $AdS$
supergroup. That is, we can write:

\begin{equation}
A=A^AT_A=\frac 12i\omega ^{ab}J_{ab}-ie^aP_a++\overline{\psi }Q  \label{ssu2}
\end{equation}
where $A$ is the gauge field of the AdS supergroup, $P_a,J_{ab},Q^\alpha $
being the generators of the AdS boosts, of the Lorentz transformations and
of the supersymmetry transformations, respectively. Hence supergravity with
cosmological constant is the gauge theory of the AdS supergroup.

The field strength associated with $A^A$ is defined as the Poincar\'{e} Lie
superalgebra-valued curvature $2$-form $R^A.$ Splitting the index $A$, we
get
\begin{equation}
\overline{R}^{ab}=R^{ab}+4\alpha ^2e^ae^b+\alpha \stackrel{\_}{\psi }\gamma
^{ab}\psi .  \label{ssu8}
\end{equation}

\begin{equation}
\stackrel{\wedge }{T\text{ }}^a=T\ ^a-\frac 12\stackrel{\_}{\psi }\gamma
^a\psi   \label{ssu9}
\end{equation}
\begin{equation}
\rho =D\psi -\alpha \gamma _a\psi e^a.  \label{ssu10}
\end{equation}

The associated Bianchi identities are given by
\begin{equation}
DR^{ab}-8\alpha ^2T\ ^ae^b+2\alpha \stackrel{\_}{\psi }\gamma ^{ab}\rho =0
\label{ssu8'}
\end{equation}
\begin{equation}
DT\ ^a+R^{ab}e_b-i\stackrel{\_}{\psi }\gamma ^a\rho =0  \label{ssu9'}
\end{equation}
\begin{equation}
D\rho -i\alpha \gamma _a\psi T^a-\frac 14R^{ab}\gamma _{ab}\psi =0.
\label{ssu10'}
\end{equation}

However, although $A^A\equiv (e^a,\omega ^{ab},\psi )$ is a Yang\_Mills
potential and $R^A\equiv (R^{ab},\stackrel{\wedge }{T\text{ }}^a,\rho )$ the
corresponding field strength,  action (\ref{ssu0}) is not of the Yang-Mills
type. The main differences between an action of the Yang-Mills type and the
action (\ref{ssu0})are: (i) a Yang-Mills action is invariant under the whole
gauge group of which the $A^A$ are the Lie superalgebra valued potentials;
(ii) the action (\ref{ssu0}), instead, is not invariant under the whole
gauge supergroup, but only under the Lorentz transformations.

The invariance under Lorentz gauge transformations is manifest. To show the
non invariance of (\ref{ssu0}) both under  a supergauge translation and
under supersymmetry we recall that, under any gauge transformation, the
gauge connection $A^A$ transforms as

\begin{equation}
\delta A=-D\lambda =d\lambda -\left[ A,\lambda \right]   \label{ssu3}
\end{equation}
with
\begin{equation}
\lambda =\frac 12i\kappa ^{ab}J_{ab}-i\rho ^aP_a++\overline{\varepsilon }Q.
\label{ssu4}
\end{equation}
Using the algebra (\ref{ssu1}) we obtain that $e^a$, $\omega
^{ab},$ and $ \psi ,$ under AdS boosts, transform as
\begin{equation}
\delta e^a=D\rho ^a;\quad \delta \omega ^{ab}=m^2(\rho ^ae^b-\rho
^be^a) \text{; \quad }\delta \psi =0;  \label{ssu5}
\end{equation}
under Lorentz rotations, as
\begin{equation}
\delta e^a=\kappa _b^ae^b;\quad \delta \omega ^{ab}=D\kappa ^{ab};\quad
\delta \psi =-\frac 12\kappa ^{ab}\gamma _{ab}\psi ;  \label{ssu6}
\end{equation}
and under supersymmetry transformations, as
\begin{equation}
\delta e^a=-2i\overline{\varepsilon }\gamma ^a\psi ;\quad \delta \omega
^{ab}=0\text{; \quad }\delta \psi =D\varepsilon .  \label{ssu7}
\end{equation}

The action (\ref{ssu0}) is invariant under diffeomorphism and under local
Lorentz rotations, but it is not invariant under neither AdS boosts
translations nor local supersymmetric transformation.

In fact, under local Poincare translations
\begin{equation}
\delta S_{AdS}=-2\int \varepsilon
_{abcd}\overline{R}^{ab}\stackrel{\wedge }{ T}^c\rho
^d+\text{surf}.\text{ term }  \label{ssu13}
\end{equation}

and under local supersymmetry transformations
\[
\delta S_{susy}=-4\int \overline{\varepsilon }\gamma _5\gamma _aD\psi
\stackrel{\wedge }{T\text{ }}^a+\text{surf. term.}
\]
Thus the invariance of the action requires the vanishing of the torsion
\begin{equation}
\stackrel{\wedge }{T\text{ }}^a=0.  \label{ssu14}
\end{equation}
This means that the connection is no longer an independent
variable. Rather, its variation is given in terms of $\delta e^a$
and $\delta \psi $, and differs from the one dictated by group
theory. The condition $\stackrel{ \wedge }{T}^a=0$ not only breaks
local Poincare invariance, but also the supersymmetry
transformations.

\section{\bf Supersymmetric extension of the Stelle-West formalism}

The basic idea of the Stelle-West $\left( SW\right) $ formalism is
founded on the mathematical definition \cite{kobaya},\cite{stelle}
of the vielbein $ V^a.$ This vielbein, also called solder form
\cite{witten}, was considered as a smooth map from the tangent
space to the space-time manifold $M$ at a point $P$ with
coordinates $x^\mu ,$ and the tangent space to the $AdS$ internal
space at the point whose $AdS$ coordinates are $\xi ^a(x)$, as the
point $P$ ranges over the whole manifold $M.$  The $fig.1$ of ref.
\cite {stelle} illustrates that such a vielbein $V_\mu ^a(x)$ is
the matrix of the map between the space $T_x(M)$ tangent to the
space-time manifold at $x^\mu , $ and the space $T_{\xi (x)}\left(
\left\{ G/H\right\} _x\right) $ tangent to the internal $AdS$
space $\left\{ G/H\right\} _x$ at the point $\xi ^a(x), $ whose
explicit form is given by $eq.(3.19)$ of ref. \cite{stelle}$.$  In
this section we consider the supersymmetric extension of the
Stelle-West formalism

\smallskip\

\subsection{\bf Non-linear realizations of supersymmetry in AdS space}

The non-linear realizations in de Sitter space can be studied by the general
method developed in ref. \cite{callan}, \cite{volkov}. Following these
references, we consider a Lie (super)group $G$ and a subgroup $H.$

Let us call $\left\{ {\bf V}_i\right\} _{i=1}^{n-d}$ the generators of $H$.
We assume that the remaining generators $\left\{ {\bf A}_l\right\} _{l=1}^d$
can be chosen so that they form a representation of $H.$  In other words,
the commutator $\left[ {\bf V}_i,{\bf A}_l\right] $ should be a linear
combination of ${\bf V}_i$ alone. A group element $g\in G$ can be
represented (uniquely) in the form
\begin{equation}
g=e^{\xi ^l{\bf A}_l}h  \label{sw1}
\end{equation}
where $h$ is an element of $H.$ The $\xi ^l$ parametrize the coset
space $ G/H.$  We do not specify here the parametrization of $h$.
One can define the effect of a group element $g_0$ on the coset
space by
\begin{equation}
g_0g=g_0(e^{\xi ^l{\bf A}_l}h)=e^{\xi ^{\prime l}{\bf A}_l}h^{\prime }
\label{sw2}
\end{equation}
or
\begin{equation}
g_0e^{\xi ^l{\bf A}_l}=e^{\xi ^{\prime l}{\bf A}_l}h_1  \label{sw3}
\end{equation}
where
\begin{equation}
h_1=h^{\prime }h^{-1}  \label{sw4}
\end{equation}
\[
\xi ^{\prime }=\xi ^{\prime }(g_0,\xi )
\]
\[
h_1=h_1(g_0,\xi ).
\]

If $g_0-1$ is infinitesimal, (\ref{sw3}) implies
\begin{equation}
e^{-\xi ^l{\bf A}_l}\left( g_0-1\right) e^{\xi ^l{\bf
A}_l}-e^{-\xi ^l{\bf A} _l}\delta e^{\xi ^l{\bf A}_l}=h_1-1.
\label{sw5}
\end{equation}
The right-hand side of (\ref{sw5}) is a generator of $H.$

Let us first consider  the case in which $g_0=h_0\in H.$ Then (\ref{sw3})
gives
\begin{equation}
e^{\xi ^{\prime l}{\bf A}_l}=h_0e^{\xi ^l{\bf A}_l}h_0^{-1}  \label{sw6}
\end{equation}
Since the $A^l$ form a representation of $H,$ this implies
\begin{equation}
h_1=h_0;\qquad h^{\prime }=h_0h.  \label{sw7}
\end{equation}

The transformation from $\xi $ to $\xi ^{\prime }$ given by (\ref{sw6}) is
linear. On the other hand, consider now
\begin{equation}
g_0=e^{\xi _0^l{\bf A}_l}.  \label{sw8}
\end{equation}
$.$ In this case (\ref{sw3}) becomes
\begin{equation}
e^{\xi _0^l{\bf A}_l}e^{\xi ^l{\bf A}_l}=e^{\xi ^{\prime l}{\bf A}_l}h.
\label{sw9}
\end{equation}
This is a non-linear inhomogeneous transformation on $\xi .$  The
infinitesimal form (\ref{sw5}) becomes
\begin{equation}
e^{-\xi ^l{\bf A}_l}\xi _0^i{\bf A}_ie^{\xi ^j{\bf A}_j}-e^{-\xi
^l{\bf A} _l}\delta e^{\xi ^i{\bf A}_i}=h_1-1.  \label{sw10}
\end{equation}
The left-hand side of this equation can be evaluated, using the algebra of
the group. Since the results must be a generator of $H$, one must set equal
to zero the coefficient of ${\bf A}_l.$ In this way one finds an equation
from which $\delta \xi ^i$ can be calculated.

The construction of a Lagrangian invariant under
coordinate-dependent group transformations requires the
introduction of a set of gauge fields $a=a_\mu ^i{\bf A}_i$d$x^\mu
,$ $\rho =\rho _\mu ^i{\bf V}_i$d$x^\mu ,$ $p=p_\mu ^l {\bf
A}_l$d$x^\mu $, $v=v_\mu ^i{\bf V}_i$d$x^\mu ,$ associated
respectively with the generators $V_i$ and $A_l$. Hence $\rho +a$
is the usual linear connection for the gauge group $G,$ and the
corresponding covariant derivatives is given by:
\begin{equation}
D=d+f(\rho +a)  \label{sw11}
\end{equation}
and its transformation law under $g\in G$ is
\begin{equation}
g:(\rho +a)\rightarrow (\rho ^{\prime }+a^{\prime })=\left[ g(\rho
+a)g^{-1}- \frac 1f(dg)g^{-1}\right]  \label{sw12}
\end{equation}
where $f$ is a constant which, as it turns out, gives the strength of the
universal coupling of the gauge fields to all other fields.

We now consider the Lie algebra valued differential form \cite{callan}
\begin{equation}
e^{-\xi ^l{\bf A}_l}\left[ d+f(\rho +a)\right] e^{\xi ^l{\bf A}_l}=p+v.
\label{sw13}
\end{equation}
The transformation laws for the forms $p(\xi ,d\xi )$ and $v(\xi ,d\xi )$
are easily obtained. In fact, using (\ref{sw8}),(\ref{sw9}) one finds \cite
{zumino}
\begin{equation}
p^{\prime }=h_1p(h_1)^{-1}  \label{sw14}
\end{equation}
\begin{equation}
v^{\prime }=h_1v(h_1)^{-1}+h_1d(h_1)^{-1}.  \label{sw15}
\end{equation}

The equation (\ref{sw14}) shows that the differential forms $p(\xi ,d\xi )$
are transformed linearly by a group element of the form (\ref{sw8}). The
transformation law is the same as by an element of $H$, except that now this
group element $h_1(\xi _0,\xi )$ is a function of the variable $\xi $.
Therefore any expression constructed with $p(\xi ,d\xi )$ which is invariant
under the subgroup $H$ will be automatically invariant under the entire
group $G$, the elements of $H$ operating linearly on $\xi $, the remaining
elements non-linearly.

We have specified the fields $p$ and $v$ as well as their transformation
properties, and now we make use of them to define the covariant derivative
with respect to the group $G$:
\begin{equation}
D=d+v.  \label{sw16}
\end{equation}
The corresponding components of the curvature two-form are
\begin{equation}
T=Dp  \label{sw17}
\end{equation}
\begin{equation}
R=dv+vv.  \label{sw18}
\end{equation}

\subsection{\bf Supersymmetric Stelle-West formalism}

We now take as $G$ the graded Lie algebra (\ref{ssu1}) having as
generators $ Q_\alpha ,P_a$ and $M_{ab}$. It has as a subalgebra
$H$ that of the de Sitter group $SO(3,2)$ with generators $P_a$
and $M_{ab}$. This, in turn, has as subalgebra $L$ that of the
Lorentz group $SO(3,1)$ with generators $ M_{ab}.$ An element of
$G$ can be represented uniquely in the form
\begin{equation}
g=e^{\overline{\chi }Q}h=e^{\overline{\chi }Q}e^{-i\xi ^aP_a}l  \label{sw19}
\end{equation}
where $h\in H$ and $l\in L.$ On can define the effect of a group element $g_0
$ on the coset space $G/H$ by
\begin{equation}
g_0g=e^{\overline{\chi }^{\prime }Q}h^{\prime }=e^{\overline{\chi }^{\prime
}Q}e^{-i\xi ^{\prime a}P_a}l^{\prime }  \label{sw20}
\end{equation}
or
\begin{equation}
g_0e^{\overline{\chi }Q}=e^{\overline{\chi }^{\prime }Q}h_1  \label{sw21}
\end{equation}
\begin{equation}
h_1e^{-i\xi ^aP_a}=e^{-i\xi ^{\prime a}P_a}l_1  \label{sw22}
\end{equation}
\begin{equation}
l_1l=l^{\prime }.  \label{sw23}
\end{equation}
Clearly $h_1=h_1(g_0,\chi )$ and $l_1=l_1(g_0,\chi ,\xi ).$

If $g_0-1$ and $h_1-1$ are infinitesimals, (\ref{sw21}),(\ref{sw22}) implies
\begin{equation}
e^{-\overline{\chi }Q}\left( g_0-1\right) e^{\overline{\chi
}Q}-e^{- \overline{\chi }Q}\delta e^{\overline{\chi }Q}=h_1-1
\label{sw24}
\end{equation}
\begin{equation}
e^{i\xi ^a{\bf P}_a}\left( h_1-1\right) e^{-i\xi ^a{\bf
P}_a}-e^{i\xi ^a{\bf P}_a}\delta e^{-i\xi ^a{\bf P}_a}=l_1-1.
\label{sw25}
\end{equation}
We consider now the following cases: If $g_0=l_0\in L,$ (\ref{sw21}),(\ref
{sw22}) give
\begin{equation}
e^{\overline{\chi }^{\prime }Q}=l_0e^{\overline{\chi }Q}l_0^{-1}
\label{sw26}
\end{equation}
\begin{equation}
h_1=l_1=l_0  \label{sw27}
\end{equation}
\begin{equation}
e^{-i\xi ^{\prime a}P_a}=l_0e^{-i\xi ^aP_a}l_0^{-1}.  \label{sw28}
\end{equation}
Both $\chi $ and $\xi $ transform linearly. If, on the other hand, we know
only that $g_0=h_0\in H,$ in particular, if
\begin{equation}
g_0=e^{-i\rho ^a{\bf P}_a}  \label{sw29}
\end{equation}
is a pseudo-translation, (\ref{sw21}) gives
\begin{equation}
e^{\overline{\chi }^{\prime }Q}=h_0e^{\overline{\chi }Q}h_0^{-1}
\label{sw30}
\end{equation}
\begin{equation}
h_1=h_0  \label{sw31}
\end{equation}
while (\ref{sw22}) gives
\begin{equation}
h_0e^{i\xi ^a{\bf P}_a}=e^{-i\xi ^{\prime a}P_a}l_1(h_0,\xi ).  \label{sw32}
\end{equation}
In this case $\chi $ transforms linearly, but the transformation law (\ref
{sw32}) of $\xi $ under pseudo-translations is inhomogeneous and non-linear.
Infinitesimally
\begin{equation}
e^{i\xi ^a{\bf P}_a}\left( -i\rho ^a{\bf P}_a\right) e^{-i\xi
^a{\bf P} _a}-e^{i\xi ^a{\bf P}_a}\delta e^{-i\xi ^a{\bf
P}_a}=l_1-1.  \label{sw33}
\end{equation}

Finally, if
\begin{equation}
g_0=e^{\overline{\varepsilon }Q}  \label{sw34}
\end{equation}
is a supersymmetry transformation, one must use (\ref{sw21}) and
(\ref{sw22} ) as they stand. Observe, however, that (\ref{sw22})
has the same form as (\ref{sw32}) except for the fact that $h_1$
is a function of $\chi $ while $ h_0$ is not. Therefore, the
transformation law of $\xi $ under a supersymmetry transformation
has the same form as that under a de Sitter transformation but,
with parameters which depend in a well defined way on $ \chi .$

An explicit form for the transformation law of $\xi ^a$ under an
infinitesimal AdS boost can be obtained from (\ref{sw33}). The result is
\begin{equation}
\delta \xi ^a=\rho ^a+\left( \frac{z\cosh z}{\sinh z}-1\right) \left( \rho
^a-\frac{\rho ^b\xi _b\xi ^a}{\xi ^2}\right)   \label{sw35}
\end{equation}
where $z=m\sqrt{(\xi ^a\xi _a)}=m\xi .$

The transformation of $\xi ^a$ under an infinitesimal Lorentz transformation
$l_0=e^{\frac i2\kappa ^{ab}J_{ab}}$ is
\begin{equation}
\delta \xi ^a=\kappa ^{ab}\xi _b  \label{sw36}
\end{equation}
and, under local supersymmetry transformation (\ref{sw34}), $\xi ^a$
transform as
\[
\delta \xi ^a=i\left( 1+\frac i6m\overline{\chi }\chi \right)
\overline{ \varepsilon }\gamma ^a\chi
\]
\[
+i\left( \frac{z\cosh z}{\sinh z}-1\right) \left( \delta _b^a-\frac{\xi
_b\xi ^a}{\xi ^2}\right) \left( 1+\frac i6m\overline{\chi }\chi \right)
\overline{\varepsilon }\gamma ^b\chi
\]
\begin{equation}
-2im\left( 1+\frac i6m\overline{\chi }\chi \right)
\overline{\varepsilon } \gamma ^{ab}\chi \xi _b.  \label{sw37}
\end{equation}

Using (\ref{sw24}) with $g_0-1=\overline{\varepsilon }Q,$ one finds that
\begin{equation}
\delta \chi =\varepsilon -\frac i6m\left( 5\overline{\chi }\chi
+\overline{ \chi }\Gamma _A\chi \Gamma ^A\right) \varepsilon
+\frac 19m^2\left( \overline{\chi }\chi \right) \varepsilon
\label{sw38}
\end{equation}
\[
h_1-1=\left( 1+\frac i6m\overline{\chi }\chi \right) \left(
\overline{ \varepsilon }\gamma ^a\chi P_a+m\overline{\varepsilon
}\gamma ^{ab}\chi J_{ab}\right) .
\]

 From (\ref{ssu2}) we know that the linear connections are given by
$ (e^a,\omega ^{ab},\psi )$. Then, based on these, we can define
the corresponding non-linear connections $(V^a,W^{ab},\Psi )$ from
(\ref{sw13}):
\[
\frac 12iW^{ab}{\bf J}_{ab}-iV^a{\bf P}_a+\overline{\Psi }Q
\]
\begin{equation}
=e^{i\xi ^a{\bf P}_a}e^{-\overline{\chi }Q}\left[ d+\frac
12i\omega ^{ab} {\bf J}_{ab}-ie^a{\bf P}_a+\overline{\psi
}Q\right] e^{\overline{\chi } Q}e^{-i\xi ^b{\bf P}_b}.
\label{sw39}
\end{equation}

The corresponding transformation laws for $V^a,W^{ab},\Psi $ can be obtained
from (\ref{sw14}),(\ref{sw15}). In fact, one can verify that, under the AdS
supergroup, the non-linear connections transform as:
\begin{equation}
\overline{\Psi }^{\prime }Q=h_1\left( \overline{\Psi }Q\right) (h_1)^{-1}
\label{sw40}
\end{equation}
\begin{equation}
-iV^{\prime a}{\bf P}_a=h_1\left( -iV^a{\bf P}_a\right) (h_1)^{-1}
\label{sw41}
\end{equation}
\begin{equation}
\frac 12iW^{\prime ab}{\bf J}_{ab}=h_1\left( \frac 12iW^{ab}{\bf
J} _{ab}\right) (h_1)^{-1}+h_1d(h_1)^{-1}.  \label{sw42}
\end{equation}

The nonlinearity of the transformation with respect to the elements of $G/H$
means that the labels associated with the parts of the algebra of $G$ which
generate $G/H$ are no longer available as symmetry indices. In other words,
the symmetry has been spontaneously broken from $G$ to $H$. An irreducible
representation of $G$ will, in general, have several irreducible pieces with
respect to $H.$ Since, in constructing invariant actions, one only needs
index saturation with respect to the subgroup $H$, as far as the invariance
is concerned it is possible to select a subset of nonlinear fields with
respect to $G$, which form irreducible multiplets with respect to $H.$

Note that, if $G=OSp(1,4)$ and $H=SO(3,1),$ the gauge fields $V^a$
form a square $4\times 4$ matrix which is invertible and can be
identified with the vierbein fields. In the same way we have that
$W^{ab}$ is a connection and $ \overline{\Psi }$ can be identified
with the Rarita-Schwinger field. These fields can be obtained from
(\ref{sw39}). The details of the calculation of $ V^a,W^{ab},\Psi
$ are given in the Appendix; the result is

\[
V^a=\Omega \left[ \cosh z\right] _{\ b}^ae^b+\Omega \left[
\frac{\sinh z}z \right] _{\ b}^a\text{D}\zeta ^b+i\left( 1-\frac
i6m\overline{\chi }\chi \right) \cdot
\]
\[
\cdot \{\left[ \overline{\chi }\gamma ^bd\chi +2\overline{\psi }\gamma
^b\chi \right] \Omega \left[ \cosh z\right] _{\ b}^a
\]
\[
-2m\xi _b\left[ \overline{\chi }\gamma ^bd\chi +2\overline{\psi }\gamma
^b\chi \right] \frac{\sinh z}z\}
\]
\[
-\frac i2\left[ 1-\frac i{12}(\overline{\chi }\gamma ^f\chi )\gamma _f+\frac
i6(\overline{\chi }\gamma ^{fg}\chi )\gamma _{fg}\right] \left( \gamma
_{cd}\omega ^{cd}-im\gamma _ce^c\right) \cdot
\]
\begin{equation}
\cdot \left[ (\overline{\chi }\gamma ^b\chi )\Omega \left[ \cosh z\right]
_{\ b}^a+2m(\overline{\chi }\gamma ^{ab}\chi )\xi _b\frac{\sinh z}z\right]
\label{sw43}
\end{equation}
where
\begin{equation}
\Omega (A)_b^a=A\delta _b^a+(1-A)\frac{\xi _b\xi ^a}{\xi ^2}  \label{sw44}
\end{equation}

\[
W^{ab}=\omega ^{ab}+m^2[\left( \xi ^ae^b-\zeta ^be^a\right) \frac{\sinh z}z
\]
\[
-m^2\left( \zeta ^aD\zeta ^b-\zeta ^bD\zeta ^a\right) \frac{\left( \cosh
z-1\right) }{z^2}]
\]
\[
-2im(1-\frac i6m\overline{\chi }\chi )\{2\overline{\psi }\gamma
^{ab}\chi +m \overline{\chi }\gamma ^{ab}d\chi
\]
\[
+m\left[ \overline{\chi }\gamma ^bd\chi +2\overline{\psi }\gamma ^b\chi
\right] \xi ^a\frac{\sinh z}z
\]
\[
+2m^2\left[ \overline{\chi }\gamma ^{cb}d\chi +2\overline{\psi }\gamma
^{cb}\chi \right] \xi ^a\xi _c\frac{\left( \cosh z-1\right) }{z^2}\}
\]
\[
+\frac 12im\left[ 1-\frac i{12}(\overline{\chi }\gamma ^f\chi
)\gamma _f+ \frac i6(\overline{\chi }\gamma ^{fg}\chi )\gamma
_{fg}\right] \left( \gamma _{cd}\omega ^{cd}-im\gamma _ce^c\right)
\cdot
\]
\begin{equation}
\cdot \left[ (\overline{\chi }\gamma ^{ab}\chi )+m(\overline{\chi
}\gamma ^a\chi )\xi ^b\frac{\sinh z}z+2m^2(\overline{\chi }\gamma
^{fb}\chi )\xi _f \frac{\left( \cosh z-1\right) }{z^2}\right]
\label{sw45}
\end{equation}

\[
\overline{\Psi }=\{\left[ 1-\frac i4m\left( 5\overline{\chi }\chi
+\overline{ \chi }\Gamma _A\chi \Gamma ^A\right) -\frac
5{24}m^2(\overline{\chi }\chi )^2\right] \overline{\psi }
\]
\[
-\frac 12[1-\frac i6m(\overline{\chi }\gamma ^a\chi )\gamma _a
\]
\[
+\frac i3m(\overline{\chi }\gamma ^{ab}\chi )\gamma _{ab}]\left( \gamma
_{cd}\omega ^{cd}-im\gamma _ce^c\right) \overline{\chi }
\]
\[
+[1-\frac i{12}m\left( 5\overline{\chi }\chi +\overline{\chi }\Gamma _A\chi
\Gamma ^A\right)
\]
\begin{equation}
-\frac 1{24}m^2(\overline{\chi }\chi )^2]d\chi \}e^{\frac 12m\xi ^d\gamma
_d}.  \label{sw46}
\end{equation}

We have specified the fields $V^a,W^{ab},$ and $\Psi $ as well as their
transformation properties, and now we make use of them to define a covariant
derivatives with respect to the group $G$:
\begin{equation}
{\cal D}=d+W.  \label{sw16}
\end{equation}
The corresponding components of a curvature two-forms are
\begin{equation}
{\cal T}^a={\cal D}V^a  \label{sw17}
\end{equation}
\begin{equation}
{\cal R}_b^a=dW_b^a+W_c^aW_b^c.  \label{sw18}
\end{equation}

\section{\bf Supergravity invariant under the AdS group}

Within the supersymmetric extension of the Stelle-West- formalism, the
action for supergravity with cosmological constant can be rewritten as

\[
S=\int \varepsilon _{abcd}{\cal R}^{ab}V^cV^d+4\overline{\Psi }\gamma
_5\gamma _a{\cal D}\Psi V^a
\]
\begin{equation}
+2\alpha ^2\varepsilon _{abcd}V^aV^bV^cV^d+3\alpha \varepsilon
_{abcd} \overline{\Psi }\gamma ^{ab}\Psi V^cV^d  \label{suads1}
\end{equation}
which is invariant under (\ref{sw40}), (\ref{sw41}), (\ref{sw42}). From such
equations we can see that the vierbein $V^a$ and the gravitino field
transform homogeneously according to the representation of the $AdS$
superalgebra but,  with the nonlinear group element $h_1\in H.$

The corresponding equations of motion are obtained by varying the
action with respect to $\xi ^a,\chi ,e^a,\omega ^{ab},\psi $. The
field equations corresponding to the variation of the action with
respect to $\xi ^a$ and $ \chi $ are not independent equations.
Following the same procedure of Ref. \cite{salga1}, we find that
equations of motion for supergravity genuinely invariant under
Super AdS are:
\begin{equation}
2\varepsilon _{abcd}\overline{{\cal R}}^{ab}V^c+4\overline{\Psi }\gamma
_5\gamma _d\rho   \label{suads5}
\end{equation}
\begin{equation}
\stackrel{\wedge }{\cal T}^{\ a}=0  \label{suads6}
\end{equation}
\begin{equation}
8\gamma _5\gamma _a\rho V^a-4\gamma _5\gamma _a\Psi
\stackrel{\wedge }{\cal T }^{\ a}=0  \label{suads7}
\end{equation}

where
\begin{equation}
\stackrel{\wedge }{\cal T}^{\ a}={\cal T}^{\ a}-\frac
i2\overline{\Psi } \gamma ^a\Psi   \label{suads8}
\end{equation}
\begin{equation}
\overline{{\cal R}}^{ab}={\cal R}^{ab}+4\alpha ^2V^aV^b+\alpha
\overline{ \Psi }\gamma ^{ab}\Psi =0  \label{suads9}
\end{equation}
\begin{equation}
\rho ={\cal D}\Psi -i\alpha \gamma ^a\Psi V^a.  \label{suads10}
\end{equation}

\subsection{{\bf Supergravity invariant under the Poincar\'{e} group}}

Taking the limit $m\rightarrow 0$ in equations (\ref{ssu1}),
(\ref{sw35}), ( \ref{sw37}), (\ref{sw38}), (\ref{sw43}),
(\ref{sw45}), (\ref{sw46}) we find that the superalgebra
(\ref{ssu1}) take the form of the superalgebra of Poincare
(\ref{su1}) and that: the transformation laws of $\xi ^a$ under an
infinitesimal Poincar\'{e} translation, under an infinitesimal
Lorentz transformation, and under local supersymmetry
transformation are given respectively by
\begin{equation}
\delta \xi ^a=\rho ^a  \label{poin0}
\end{equation}
\begin{equation}
\delta \xi ^a=\kappa ^{ab}\xi _b  \label{poin1}
\end{equation}
\begin{equation}
\delta \xi ^a=i\overline{\varepsilon }\gamma ^a\chi ;  \label{poinc1}
\end{equation}
the transformation laws of $\chi $ under an infinitesimal Poincar\'{e}
translation, under an infinitesimal Lorentz transformation, and under local
supersymmetry transformation are given respectively by
\begin{equation}
\delta \chi =0  \label{poin2}
\end{equation}
\begin{equation}
\delta \chi =0  \label{poin3}
\end{equation}
\begin{equation}
\delta \chi =\varepsilon .  \label{poin4}
\end{equation}

In this limit $G=ISO(3,1)$ and $H=SO(3,1)$ and the fields vierbein $V^a,$
the connection $W^{ab}$ and the Rarita-Schwinger field $\overline{\Psi }$
are given by

\begin{equation}
V^a=e^a+D\zeta ^a+i\left( 2\overline{\psi }+D\overline{\chi }\right) \gamma
^a\chi  \label{poin5}
\end{equation}

\begin{equation}
W^{ab}=\omega ^{ab}  \label{poin6}
\end{equation}

\begin{equation}
\overline{\Psi }=\overline{\psi }+D\overline{\chi }  \label{poin7}
\end{equation}
where now

\begin{equation}
D=d+\omega .  \label{poin8}
\end{equation}
The corresponding components of the curvature two-form are now
\begin{equation}
{\cal T}^{\ a}=DV^a  \label{poin9}
\end{equation}
\begin{equation}
R_b^a=d\omega _b^a+\omega _c^a\omega _b^c.  \label{poin10}
\end{equation}

The limit $m\rightarrow 0$ of the action \ref{suads1} is obviously the
action for $N=1$ Supergravity in $(3+1)$-dimensions:
\begin{equation}
S=\int \varepsilon _{abcd}R^{ab}V^cV^d+4\overline{\Psi }\gamma _5\gamma
_aD\Psi V^a  \label{poin11}
\end{equation}
which is genuinely invariant under the Poincar\'{e} supergroup. In fact, it
is direct to verify that the action (\ref{poin11}) is invariant under (\ref
{poin0}-\ref{poin4}) plus the transformation law of $e^a,\omega ^{ab},\psi $
under infinitesimal Poincar\'{e} translations, under infinitesimal Lorentz
transformations, and under local supersymmetry transformations, which are
given by

\[
\delta \omega ^{ab}=-D\kappa ^{ab};\delta e^a=\kappa _{\text{ }
b}^ae^b;\delta \psi =\frac 14\kappa ^{ab}\gamma _{ab}\psi
\]

\[
\delta \omega ^{ab}=0;\delta e^a=D\rho ^a;\delta \psi =0;\delta \xi ^a=-\rho
^a
\]

\[
\delta \omega ^{ab}=0;\delta e^a=i\overline{\varepsilon }\gamma ^a\psi
;\delta \psi =D\varepsilon ;\delta \xi ^a=0.
\]

The corresponding field equations are given by
\begin{equation}
2\varepsilon _{abcd}R^{ab}V^c+4\overline{\Psi }\gamma _5\gamma _dD\Psi
\label{suads5}
\end{equation}
\begin{equation}
\stackrel{\wedge }{\cal T}^{\ a}=0  \label{suads6}
\end{equation}
\begin{equation}
8\gamma _5\gamma _aD\Psi V^a-4\gamma _5\gamma _a\Psi
\stackrel{\wedge }{\cal T}^{\ a}=0  \label{suads7}
\end{equation}

where
\begin{equation}
\stackrel{\wedge }{\cal T}^{\ a}={\cal T}^{\ a}-\frac
i2\overline{\Psi } \gamma ^a\Psi .  \label{suads8}
\end{equation}

In ref. \cite{salga} we have clamed that the successful formalism used by
Stelle-West and by Grignani-Nardelli \cite{Grigna} to construct an action
for $(3+1)$-dimensional gravity invariant under the Poincar\'{e} group can
be generalized to supergravity in $(3+1)$-dimensions. In fact, that is
correct. Using the vierbein of Stelle-West and Grignani-Nardelli, one gets a
supergravity action invariant under Poincare translation. However the action
of ref. \cite{salga} is not invariant under supersymmetry transformations as
we can see from (\ref{su13}).

To obtain an action invariant both under  Poincare translations and under
supersymmetry transformations, we must carry out the supersymmetric
extension of the Stelle-West formalism. The correct vierbein, spin
connection, and gravitino field to construct a supergravity action (see \ref
{poin11}) genuinely invariant under the Poincar\'{e} supergroup are given in
equations (\ref{poin5}), (\ref{poin6}), (\ref{poin7}).

\section{\bf Comments and possible developments}

The main results of this work can be summarized as follows:

$(i)$ In order to construct a gauge theory of the supersymmetric extension
of the $AdS$ group, it is necessary to carry out the supersymmetric
extension of the Stelle-West-Grignani-Nardelli formalism.

$(ii)$ The correspondence with the usual $N=1$ supergravity with
cosmological constant formulation has been established by giving the
expressions, in terms of the gauge fields, of the spin connection, the
vierbein, and the gravitino. These fields are given by complicated
expressions involving $\xi ^a,\chi ,\psi ,\omega ^{ab}$ and $e^a$.

$(iii)$ An action for $(3+1)$-dimensional $N=1$ supergravity with
cosmological constant genuinely invariant under the supersymmetric extension
of the $AdS$ group has been proposed. The corresponding equations of motion
reproduce the usual equation for $N=1$ supergravity with cosmological
constant.

Several aspects deserve consideration and many possible developments can be
worked out. An old and still unsolved problem is the construction of an
eleven dimensional supergravity off-shell invariant under the supersymmetric
extension of the $AdS$ group (work in progress). The construction of an
action for supergravity in ten dimensions genuinelly invariant under the $AdS
$ superalgebra, and its relation to eleven dimensional supergravity, could
also be of interest.

Another interesting issue is the connection between the present paper and
the supergravity in $(3+1)$-dimensions obtained via dimensional reduction
from five-dimensional Chern-Simons supergravity (work in progress).

\smallskip\

\smallskip\

\section{\bf Appendix}

In this appendix, we discuss how to derive some of the results given in the
text, in particular the expressions for $V^a,W^{ab},\Psi .$  We use the
techniques of refs. \cite{stelle},\cite{zumino}, which we summarize here for
convenience.

For any two quantities $X$ and $Y$ we define
\begin{equation}
\left[ X,Y\right] \equiv X\wedge Y  \label{ap1}
\end{equation}
\begin{equation}
X^2\wedge Y=\left[ X,\left[ X,Y\right] \right] .  \label{ap2}
\end{equation}

\begin{itemize}
\item  The expression $f(X)\wedge Y$ is defined as a series of
multiple commutators, obtained by expanding the function $f(X)$ as
a power series in $ X.$  It is direct to verify that
\begin{equation}
g(X)\wedge \left[ f(X)\wedge Y\right] =\left[ g(X)f(X)\right] \wedge Y.
\label{ap3}
\end{equation}
\end{itemize}

As a consequence, the equation
\begin{equation}
f\left( X\right) \wedge Y=Z  \label{ap4}
\end{equation}
can be solved for $Y$ in the form
\begin{equation}
Y=\left[ f\left( X\right) \right] ^{-1}\wedge Z.  \label{ap5}
\end{equation}
In particular, we have
\begin{equation}
e^XYe^{-X}=e^X\wedge Y  \label{ap6}
\end{equation}
\begin{equation}
e^X\delta e^{-X}=\frac{1-e^X}X\wedge \delta X  \label{ap7}
\end{equation}
where $\delta $ is any variation.

When written in the above notation, eq. (\ref{sw33}) become
\begin{equation}
e^{i\xi ^a{\bf P}_a}\wedge \left( -i\rho ^a{\bf P}_a\right) -\frac{1-e^{i\xi
^a{\bf P}_a}}{i\xi ^b{\bf P}_b}\wedge \left( i\delta \xi ^a{\bf P}_a\right)
=l_1-1.  \label{ap8}
\end{equation}

Since this is a Lorentz generator, we must evaluate the AdS boost component
of the left-hand side and set it equal to zero; only commutators of even
order contribute to it.\ Therefore we must take the even powers of $i\delta
\xi ^a{\bf P}_a$ of the functions occurring in (\ref{ap8}). This leads to
\begin{equation}
\delta \xi ^aP_a=\rho ^aP_a+\left( \frac{z\cosh z}{\sinh z}-1\right) \left(
\rho ^a-\frac{\rho ^b\xi _b\xi ^a}{\xi ^2}\right) P_a.  \label{ap9}
\end{equation}

In a similar way, we can make use of (\ref{sw24}) with
$g_0-1=\overline{ \varepsilon }Q;$ one finds that
\begin{equation}
e^{-\overline{\chi }Q}\wedge \left( \overline{\varepsilon
}Q\right) -\frac{ 1-e^{-\overline{\chi }Q}}{^{\overline{\chi
}Q}}\wedge \left( \delta \overline{\chi }Q\right) =h_1-1.
\label{ap10}
\end{equation}

Here there is a simplification: any power of $\chi $ higher than four
vanishes identically due to the anticommuting property of the spinor
component. Therefore we need only
\begin{equation}
\overline{\chi }Q\wedge \overline{\varepsilon }Q=-2m\overline{\chi }\gamma
^{AB}\varepsilon J_{AB}  \label{ap11}
\end{equation}
\begin{equation}
(\overline{\chi }Q)^2\wedge \overline{\varepsilon }Q=-\frac
52im\overline{ \chi }\chi \overline{\varepsilon }Q-\frac
12im\overline{\chi }\Gamma _A\chi \overline{\varepsilon }\Gamma
^AQ  \label{ap12}
\end{equation}
\begin{equation}
(\overline{\chi }Q)^3\wedge \overline{\varepsilon
}Q=4im^2\overline{\chi } \chi \overline{\chi }\gamma
^{AB}\varepsilon J_{AB}  \label{ap13}
\end{equation}
\begin{equation}
(\overline{\chi }Q)^4\wedge \overline{\varepsilon
}Q=-5m^2(\overline{\chi } \chi )^2\overline{\varepsilon }Q,
\label{ap14}
\end{equation}

where the five matrices
\[
\Gamma ^A\equiv \left( \gamma _a\gamma _5,\gamma _5\right)
\]
satisfy
\[
\Gamma _A\Gamma _B+\Gamma _B\Gamma _A=2\eta _{AB}
\]
\[
\gamma _{AB}=\frac 14\left[ \Gamma _A,\Gamma _B\right]
\]
\[
2m\gamma ^{AB}J_{AB}=2\gamma ^aP_a-2m\gamma ^{ab}J_{ab}
\]

If one sets equal zero, in the left-hand side of (\ref{ap10}), the part with
the even powers of $\overline{\chi }Q$, one finds
\begin{equation}
\cosh \left( \overline{\chi }Q\right) \wedge \overline{\varepsilon
}Q-\frac{ \sinh \left( \overline{\chi }Q\right) }{\overline{\chi
}Q}\wedge \delta \overline{\chi }Q=0.  \label{ap15}
\end{equation}
Using (\ref{ap4}), we have
\begin{equation}
\delta \overline{\chi }Q=\left[ 1+\frac 13(\overline{\chi
}Q)^2-\frac 1{45}( \overline{\chi }Q)^4\right] \wedge
\overline{\varepsilon }Q.  \label{ap16}
\end{equation}
If one now makes use of (\ref{ap11}) to (\ref{ap14}), one obtains

\begin{equation}
\delta \chi Q=\left[ \varepsilon -\frac i6m\left( 5\overline{\chi
}\chi + \overline{\chi }\Gamma _A\chi \Gamma ^A\right) \varepsilon
+\frac 19 m^2\left( \overline{\chi }\chi \right) ^2\varepsilon
\right] Q.  \label{ap17}
\end{equation}
On the other hand, using (\ref{ap16}), the part with the odd powers gives
\[
h_1-1=\left( 1+\frac i6m\overline{\chi }\chi \right) \left(
\overline{ \varepsilon }\gamma ^a\chi P_a+m\overline{\varepsilon
}\gamma ^{ab}\chi J_{ab}\right) .
\]

The nonlinear fields $V^a,W^{ab},\Psi $ are evaluated from their
definition ( \ref{sw13}), (\ref{sw39}) following the same
procedure of ref. \cite{stelle}.

{\bf Acknowlegments}
(P.S) wishes to thank Herman Nicolai for the
hospitality in AEI in Golm bei Potsdam and Deutscher Akademischer
Austauschdienst (DAAD) for financial support. SdC wishes to thank
F. M\"{u}ller-Hoissen for the hospitality in G{ \"{o}}ttingen
where part of this work was dones, and MECESUP FSM 9901 grant for
financial support. The support from grants FONDECYT Projects N$^0$
1000305 (SdC) and 1010485 (SdC and PS), and by grants UCV-DGIP
N$^0$ 123.752 (SdC), UdC/DI N$^0$ 202.011.031-1.0 (PS) are also
acknowledged. The authors are grateful to Universidad de
Concepci\'{o}n for partial support of the 2$ ^{nd}$ Dichato
Cosmological Meeting, where this work was started.

\end{document}